\documentclass[journal=jpclcd,manuscript=letter,layout=twocolumn]{achemso}
\usepackage{notes2bib}
\usepackage{graphicx}

\title{Threshold Energies for Single Carbon Knockout from Polycyclic Aromatic Hydrocarbons}

\author{M.~H.~Stockett}
\affiliation[Stockholm University]{Department of Physics, Stockholm University, Stockholm, SE-106 91, Sweden}
\alsoaffiliation[Aarhus University]{Department of Physics and Astronomy, Aarhus University, DK-8000 Aarhus C, Denmark}
\email{stockett@phys.au.dk}
\author{M.~Gatchell}
\affiliation[Stockholm University]{Department of Physics, Stockholm University, Stockholm, SE-106 91, Sweden}
\author{T.~Chen}
\affiliation[Stockholm University]{Department of Physics, Stockholm University, Stockholm, SE-106 91, Sweden}
\author{N.~de~Ruette}
\affiliation[Stockholm University]{Department of Physics, Stockholm University, Stockholm, SE-106 91, Sweden}
\author{L.~Giacomozzi}
\affiliation[Stockholm University]{Department of Physics, Stockholm University, Stockholm, SE-106 91, Sweden}
\author{M.~Wolf}
\affiliation[Stockholm University]{Department of Physics, Stockholm University, Stockholm, SE-106 91, Sweden}
\author{H.~T.~Schmidt}
\affiliation[Stockholm University]{Department of Physics, Stockholm University, Stockholm, SE-106 91, Sweden}
\author{H.~Zettergren}
\affiliation[Stockholm University]{Department of Physics, Stockholm University, Stockholm, SE-106 91, Sweden}
\author{H.~Cederquist}
\affiliation[Stockholm University]{Department of Physics, Stockholm University, Stockholm, SE-106 91, Sweden}

\date{\today}

\begin{document}

\maketitle

\begin{tocentry}
\includegraphics[width=0.98\columnwidth]{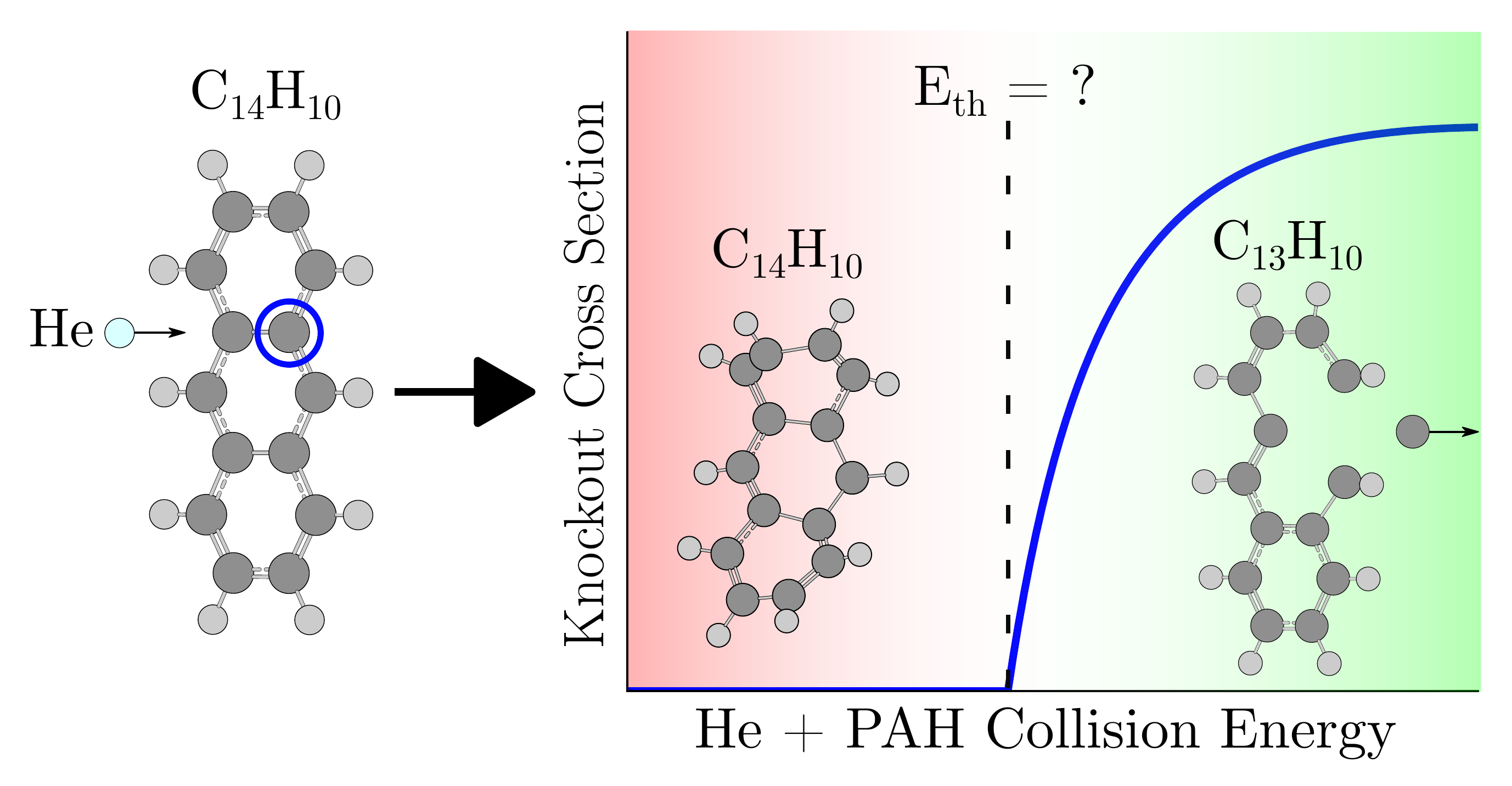}
\end{tocentry}

\begin{abstract}
We have measured absolute cross sections for ultrafast (fs) single-carbon knockout from Polycyclic Aromatic Hydrocarbon (PAH) cations as functions of He-PAH center-of-mass collision energy in the range 10-200 eV. Classical Molecular Dynamics (MD) simulations cover this range and extend up to 10$^5$ eV. The shapes of the knockout cross sections are well described by a simple analytical expression yielding experimental and MD threshold energies of $E_{th}^{Exp}=32.5\pm 0.4$ eV and $E_{th}^{MD}=41.0\pm 0.3$ eV, respectively. These are the first measurements of knockout threshold energies for molecules isolated \emph{in vacuo}. We further deduce semi-empirical (SE) and MD displacement energies --- \emph{i.e.} the energy transfers to the PAH molecules at the threshold energies for knockout --- of $T_{disp}^{SE}=23.3\pm 0.3$ eV and $T_{disp}^{MD}=27.0\pm 0.3$ eV. The semi-empirical results compare favorably with measured displacement energies for graphene $T_{disp}=23.6$ eV [Meyer \emph{et al.} Phys. Rev Lett. \textbf{108} 196102 (2012) and \textbf{110} 239902 (2013)].
\end{abstract}

Polycyclic Aromatic Hydrocarbons (PAHs) are suggested to be ubiquitous throughout the interstellar medium \cite{tielens13}.  A strong indication for this is the observation of infrared emission bands at wavelengths associated with the vibrational modes of fused hexagonal benzene ring structures which are typical of PAH molecules.  These so-called Aromatic Infrared Bands are however very similar for different PAH species and can not be used for unique identification of specific PAHs.  It is usually argued that interstellar PAHs must be large, containing over 50 C atoms \cite{tielens08}, as smaller molecules would be destroyed through dissociation induced by absorption of UV-radiation.  Recent studies \cite{martin13}, however, have shown that small PAHs such as anthracene (C$_{14}$H$_{10}$) may radiatively cool more rapidly than previously thought, possibly increasing their survival probabilities in the ISM.

Another important destruction mechanism for PAHs --- large or small --- is collisions with fast atoms or ions, particularly H and He.  Such collisions, with center-of-mass energies in the 10-1000 eV range, should be especially relevant when shockwaves from exploding supernovae interact with regions containing PAHs \cite{micelotta2010}, or when stellar winds interact with atmospheres such as that of Titan \cite{lopezpuertas13,waite07}.  It was recently shown that the direct, prompt (fs) knockout of single C or H atoms is an important mechanism for PAH destruction under these conditions, and that this may be the dominant channel for PAHs containing more than about 20 C atoms \cite{stockett14}.  The clearest experimental signature of this process is the detection of charged fragments from which a single carbon atom (perhaps accompanied by one or several H atoms) has been lost from the parent molecule:

\begin{equation}
\textrm{C$_n$H$_m$\textsuperscript{+}+He$\longrightarrow$C$_{n-1}$H$_{m-x}$\textsuperscript{+}+CH$_x$+He ($x=0$, 1,\ldots).}
\label{eq_ko}
\end{equation}

Such single-carbon loss processes are highly disfavored in statistical fragmentation processes, in which internal excitation energies are distributed over all degrees of freedom prior to fragmentation and where the lowest-energy dissociation channels dominate.  For PAHs, H- and C$_2$H$_2$-loss are typical statistical fragmentation channels governed by dissociation energies in the range of 5-7 eV \cite{holm11}.  The energy required to remove a single C atom is much higher due to substantial dissociation barriers (17 eV in the case of coronene \cite{stockett14}). The CH$_x$-loss channel is therefore usually not activated when the internal energy is distributed over all degrees of freedom. This is most often the case when the PAHs are excited by photons \cite{brechignac99,pirali06,rapacioli05AA,reitsma14}, electrons \cite{denifl05,denifl06} or in collisions with ions at several keV of energy or more \cite{postma10,mishra14,reitsma13,chen14,holm10}. All of these activation agents primarily excite the molecule's electrons, followed by fast (ps) internal vibrational redistribution and delayed ($\mu$s) fragmentation.  However, in collisions with ions, atoms or other heavy particles at lower collision energies, nuclear stopping processes --- Rutherford-like scattering of the projectile ion or atom on individual (screened) atomic nuclei in the molecules --- becomes important \cite{stockett14}.  The energy transfer to these nuclei may then be sufficient to directly remove them, \emph{i.e.} to knock them out, on femtosecond (fs) timescales.  In this scenario, fragmentation occurs before the energy is distributed over the internal degrees of freedom. From this point of view the fragmentation of the molecule is \textit{not} statistical while the distributions of impact parameters and molecular orientations are inherently statistical.  

For PAHs, single-carbon knockout may result in fragments with pentagons and non-planar structures \cite{stockett14}.  These features may be important in understanding the astrophysical formation of fullerenes such as C$_{60}$, which have been observed in numerous interstellar environments \cite{cami10,roberts12,campbell15}.  Several groups have investigated possible ``top-down'' fullerene formation routes from graphene flakes \cite{chuvilin10}, amorphous carbon clusters \cite{micelotta12}, and large PAHs \cite{berne12,zhen14}, but have so far not discussed carbon knockout in collisions with fast ions or atoms in this context. 

Knockout fragmentation of C$_{60}$ was first reported by Larsen \emph{et al.} \cite{larsen99}, with additional work carried out by Tomita \emph{et al.} \cite{tomita02}.  In both experiments \cite{larsen99,tomita02}, low abundances of C$_{59}$\textsuperscript{+} were produced in collisions between He or Ne atoms and C$_{60}$\textsuperscript{-}.  More recently, Gatchell \emph{et al.} reported the detection of C$_{59}$\textsuperscript{+} formed in C$_{60}$\textsuperscript{+}+Ne collisions \cite{gatchell14b}.  Formation of C$_{59}$\textsuperscript{+} has been shown to be the first step in the formation of covalently bound dumbbell-shaped C$_{119}$\textsuperscript{+} inside clusters of fullerene molecules bombarded by keV ions \cite{zettergren13,seitz13,wang14,gatchell14}.  

These earlier studies with PAHs and C$_{60}$ have all emphasized that experimental detection of knockout fragments depends sensitively on the internal energy remaining in the large molecular fragment (see Equation \ref{eq_ko}). Close collisions leading to single-carbon knockout often lead to internal energies suffcient for secondary statistical fragmentation \cite{stockett14b} --- depleting the unique experimental signal (C-loss for fullerenes; CH$_x$-loss for PAHs) for carbon knockout. In this Letter, we report experimental absolute cross sections for single carbon knockout for collisions between He atoms and three PAH cations of different mass: anthracene C$_{14}$H$_{10}$, pyrene C$_{16}$H$_{10}$, and coronene C$_{24}$H$_{12}$, as functions of the center-of-mass energy, $E_{CM}$, in the He-PAH$^+$ system. In addition, we present classical Molecular Dynamics (MD) simulations for these processes, where we have used the Tersoff potential \cite{PhysRevB.37.6991,PhysRevB.39.5566} to describe the breaking and forming of bonds within the PAH molecules and the ZBL potential \cite{zbl_pot_book} for the interactions between the He atom and all individual atoms in the molecule. Below, we fit an analytical expression \cite{chen14} for the relation between the knockout cross section, the He-PAH$^+$ center-of-mass energy, E$_{CM}$, and the threshold energy in this frame of reference, $E_{th}$. We deduce individual threshold energies from the experimental data, $E_{th}^{Exp}$, and our MD-simulations, $E_{th}^{MD}$.

The microscopic quantity which gives rise to the observed threshold is known as the displacement energy $T_{disp}$. This is the energy transferred to the molecular system at threshold. The value of $T_{disp}$ for PAHs has been the subject of debate. Postma \emph{et al.} recently reported values of $T_{disp}$ for PAHs around 27 eV \cite{postma14} based on classical MD simulations for specific orientations of the molecule in collisions with He. These calculations provided more accurate values than those estimated in earlier work \cite{micelotta2010}. However, the results of Ref. \cite{postma14} is significantly higher (by about 20$\%$) than those from theoretical and experimental studies of electron and heavy ion bombardment of graphene monolayers \cite{zobelli07,kotakoski10,lehtinen10,meyer12,meyer13}. The reason for this difference is an interesting open question, as PAHs resemble small graphene flakes with similar carbon ring structures. Here, we report a semi-empirical value of the displacement energy ($T_{disp}^{SE}=23.3\pm 0.3$ eV) based on energy transfers extracted from our MD simulations and our experimentally measured threshold energies. Our result for PAHs is close to the value found for electron impact on graphene (23.6 eV) \cite{meyer12,meyer13}, suggesting that the displacement energies are similar for PAHs and graphene.

\begin{figure}
\centering
\includegraphics[width=0.98\columnwidth]{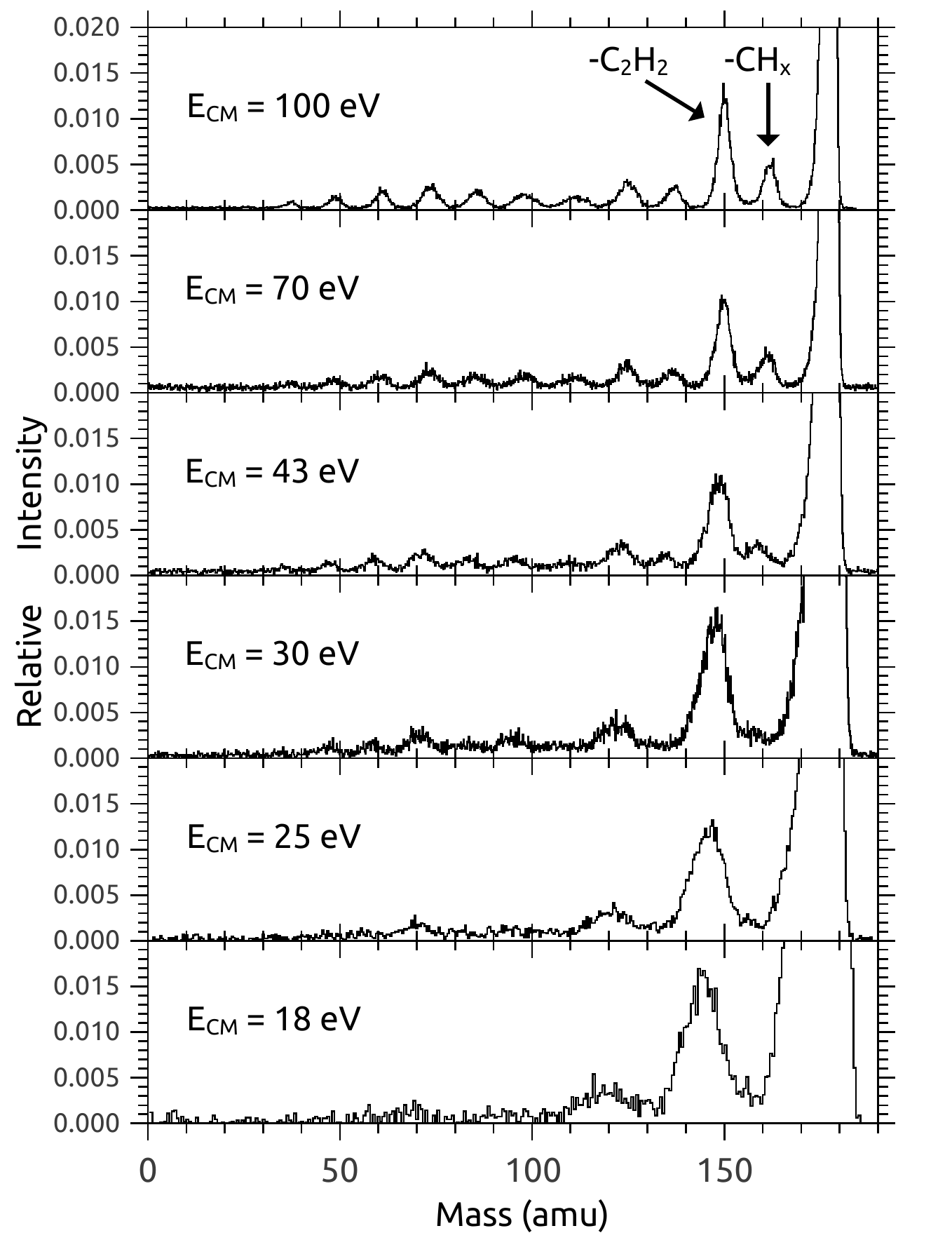}
\caption{Measured mass spectra following C$_{14}$H$_{10}$\textsuperscript{+}+He collisions at several values of the He-PAH$^+$ center-of-mass energy $E_{CM}$.}
\label{fig_anthspectra}
\end{figure}

In Figure \ref{fig_anthspectra}, we show mass spectra for collisions between anthracene cations (C$_{14}$H$_{10}$\textsuperscript{+}) and He at different He-PAH$^+$ center-of-mass energies $E_{CM}$. The plots are normalized such that the height of the parent peak (unfragmented C$_{14}$H$_{10}$\textsuperscript{+}) is unity. These spectra are measured using an electrostatic energy analyzer. Peaks for fragments with different numbers of H atoms are not resolved due to the distributions of kinetic energies of the fragments. The top frame, showing the spectrum taken at $E_{CM}=100$ eV is representative of all the measurements above 100 eV.  The fragment peak with the highest mass, corresponding to the knockout of a single carbon atom, decreases in intensity as the center-of-mass energy is decreased, until it becomes barely visible in the $E_{CM}=30$ eV spectrum. At the lowest energies ($E_{CM}<30$ eV), Gaussian fits to the CH$_x$-loss and neighbouring peaks yield insignificant CH$_x$-loss cross sections (see the Supporting Information). The shifts of the fragment peaks towards lower apparent masses as E$_{CM}$ is lowered are due to larger fractions of the available energy being converted to internal energies of the molecular systems. The internal energy remaining in the C$_{n-1}$H$_{m-x}$\textsuperscript{+} fragments after knockout is important, as it determines their probability to survive against secondary statistical fragmentation on the experimental timescale (some tens of $\mu$s).

\begin{figure}
\centering
\includegraphics[width=0.98\columnwidth]{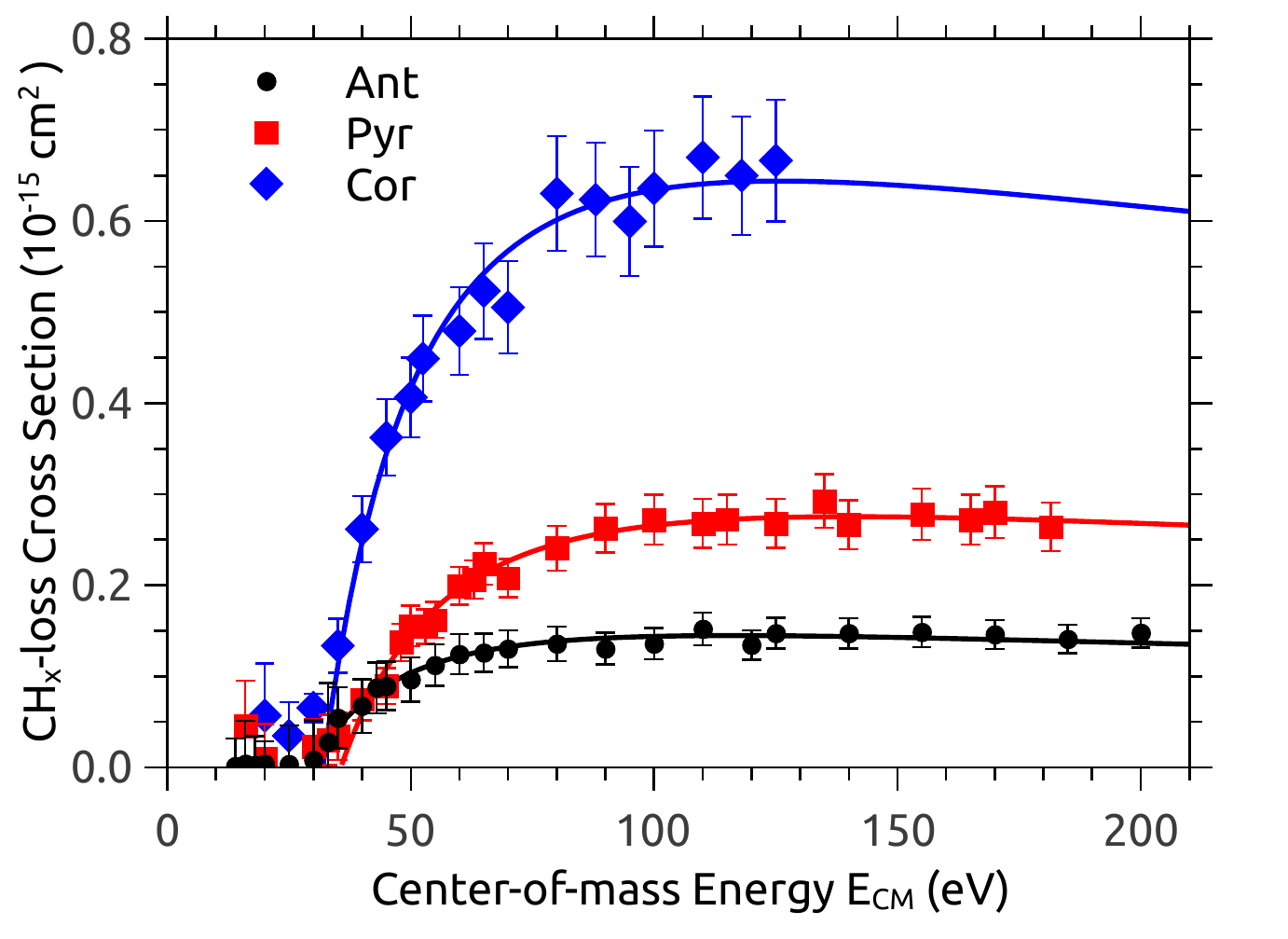}
\caption{Experimental absolute cross sections for CH$_x$-loss in C$_{14}$H$_{10}$\textsuperscript{+}+He, C$_{16}$H$_{10}$							\textsuperscript{+}+He, and C$_{24}$H$_{12}$\textsuperscript{+}+He collisions as functions of the He-PAH$^+$ center-of-mass energy $E_{CM}$. Solid lines are fits to Equation \ref{eq_tao} where the threshold energies $E_{th}$ and amplitude factors $A$ have been allowed to vary individually for the three PAH molecules.}
\label{fig_comparepahs}
\end{figure}

The threshold behavior may be clearly seen in Figure \ref{fig_comparepahs} where we plot the absolute experimental cross sections for CH$_x$-loss as functions of the center-of-mass energy, $E_{CM}$, for anthracene C$_{14}$H$_{10}$\textsuperscript{+}, pyrene C$_{16}$H$_{10}$\textsuperscript{+}, and coronene C$_{24}$H$_{12}$\textsuperscript{+}, colliding with He. These cross sections are simply the products of the measured total fragmentation cross sections and the relative intensities of the CH$_x$-loss peaks. For each PAH ion in Figure \ref{fig_comparepahs}, a rapid increase in the CH$_x$-loss cross section is observed between $\approx 30$ and $\approx 50$ eV, followed by a plateau region at higher center-of-mass collision energies. 

Chen \emph{et al.} \cite{chen14} give an analytical expression for the cross section $\sigma_{He-C}$ for energy transfer greater than some threshold energy $E_{th}^{He-C}$ in binary collisions between He and C atoms: 

\begin{equation}
\sigma_{He-C}=\frac{A^{He-C}/E_{CM}^{He-C}}{\pi^2 \arccos^{-2} (\sqrt{E_{th}^{He-C}/E_{CM}^{He-C}})-4}.
\label{eq_taohec}
\end{equation}

\noindent where $A^{He-C}$ is a constant \cite{chen14}. The He-C center-of-mass collision energy is $E_{CM}^{He-C}=\frac{M_C}{M_{He}+M_C}E_{He}$, where $M_C$ and $M_{He}$ are the masses of the colliding atoms and $E_{He}$ is the kinetic energy of the He projectile in the rest frame of the C target. In Ref. \cite{chen14}, this expression was used to calculate absolute cross sections for carbon knockout from PAHs by assuming that the energy transfer to the C atom had to be at least 27 eV \cite{postma14} and by scaling $\sigma_{He-C}$ by the number of carbon atoms $N$ in the PAH (\emph{i.e.} $\sigma_{KO}=N\sigma_{He-C}$). Here we will instead express the knockout (KO) cross section in terms of the center-of-mass energy in the He-PAH$^+$ system, $E_{CM}$. As PAHs are much heavier than He, the collision energy in this system is very close to the kinetic energy of the He atom in the PAH-target frame. Thus $E_{CM}\simeq E_{He}=\frac{4}{3}E_{CM}^{He-C}$. As mentioned above, the threshold energy $E_{th}$ refers to the smallest energy for knockout in the He-PAH$^+$ center-of-mass system and is then $E_{th}=\frac{4}{3}E_{th}^{He-C}$. This yields

\begin{equation}
\sigma_{KO}=\frac{A/E_{CM}}{\pi^2 \arccos^{-2} (\sqrt{E_{th}/E_{CM}})-4}
\label{eq_tao}
\end{equation}

\noindent where $A=\frac{4}{3}NA^{He-C}$ has the value: 

\begin{equation}
A=\frac{4}{3}0.736N\frac{2\pi a_0Z_{He}Z_C}{\sqrt{Z_{He}^{2/3}+Z_C^{2/3}}}\frac{4\pi \epsilon_0}{e^2}.
\label{eq_A}
\end{equation}

\noindent Here $a_0$ is the Bohr radius, $Z_{He}$ and $Z_C$ are the atomic numbers of He and C, $\epsilon_0$ is the vacuum permittivity and $e$ is the electron charge. The solid lines in Figure \ref{fig_comparepahs} are fits to the experimental cross sections where $E_{th}$ and $A$ have been allowed to vary. These fits give threshold energies $E_{th}^{Exp}$ of $29.4\pm 0.3$, $35.6\pm 0.4$ and $32.0\pm 0.6$ eV for anthracene, pyrene, and coronene, respectively. The weighted average of these three individual threshold values, using the estimated errors as weights, is $\langle E_{th}^{Exp}\rangle=32.5\pm 0.4$ eV. 

\begin{figure}
\centering
\includegraphics[width=0.98\columnwidth]{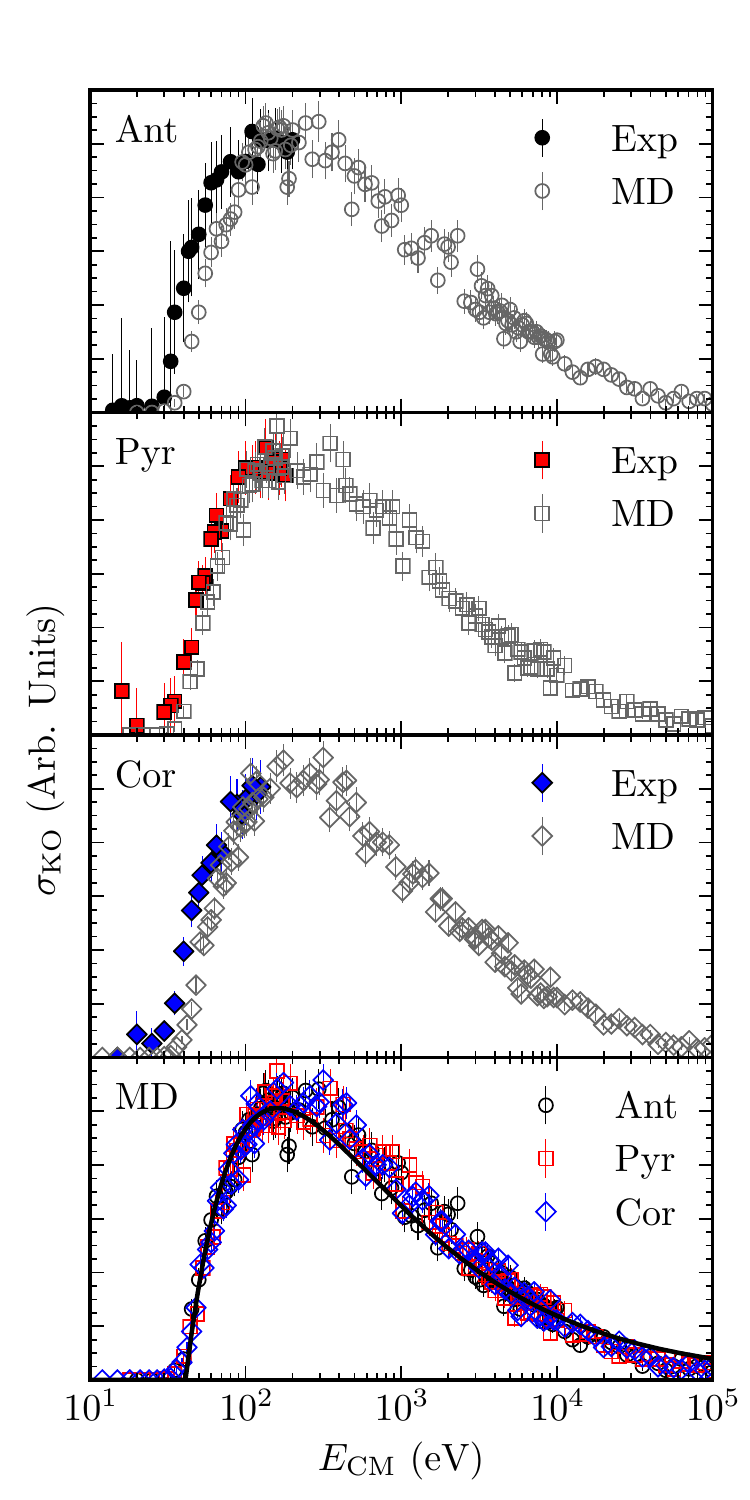}
\caption{Classical MD (Tersoff) simulated (open symbols) and experimental (filled symbols) cross sections for CH$_x$-loss for anthracene (Ant), pyrene (Pyr) and coronene (Cor) in collisions with He as functions of the center-of-mass energy $E_{CM}$. The bottom panel shows normalized results from the simulations for all three molecules.  The solid line is a fit to Equation \ref{eq_tao} for this combined dataset.}
\label{fig_md}
\end{figure}

Figure \ref{fig_md} shows the relative cross sections for single carbon knockout obtained from our MD simulations for all three PAHs over a broad energy range, along with the experimental data. The peak values of all the distributions are normalized to 1.0 (based on the average in the plateau region). There is good agreement in the shapes of the experimental and MD distributions. The MD results are well-represented by Equation \ref{eq_tao} for energies up to at least 10$^5$ eV. The bottom panel shows a fit to Equation \ref{eq_tao} for the combined normalized MD dataset for all three molecules, yielding an average threshold value for knockout of $\langle E_{th}^{MD}\rangle =41.0\pm 0.3$ eV. 

\begin{table}
\centering
\begin{tabular}{c|cccc}
 & Eq. \ref{eq_A} & MD & MD & Exp. \\

PAH &  &(S+D)&(S Only)&(S Only)\\ 
\hline 
Ant& 358 & 267 & 243 & 85 \\ 
 
Pyr& 408 & 316 & 287 & 196 \\ 

Cor& 612 & 458 & 400 & 411 \\ 
\end{tabular} 
\caption{Amplitude factors $A$ (in units of 10$^{-15}$ cm$^{2}\cdot$eV) from Equation \ref{eq_A}, from fits to the MD simulated cross sections for single C knockout (S only), single and double C knockout (S+D) (see Supporting Information), and from fits to the present experimental cross sections.}
\label{tab_A}
\end{table}

The values of $A$ from Equation \ref{eq_A} are given in Table \ref{tab_A}, along with those obtained from fits to the \emph{absolute} cross sections for the individual PAHs from our MD simulations (see Supporting Information) and from the present measurements (Figure \ref{fig_comparepahs}). Also given in Table \ref{tab_A} are the MD cross sections when we include the contributions from double-carbon knockout processes (double-knockout processes are due to single interactions with two different C-atoms in the molecule and secondary processes where the first carbon atom knocks out  another carbon atom, see Supporting Information). The fits to the experimental cross sections give $A$-values which are significantly lower than the results of the simulations and Equation \ref{eq_A}, particularly for anthracene and pyrene. This reflects the survival probabilities for the C$_{n-1}$H$_{m-x}$\textsuperscript{+} fragments produced in the collision (Equation \ref{eq_ko}), which are not included in the simulations or Equation \ref{eq_tao}. Larger PAHs such as coronene have more internal degrees of freedom over which to distribute the energy deposited in the collision, and are thus less likely to undergo secondary statistical fragmentation following knockout on the present experimental time scale of tens of microseconds \cite{stockett14}. The experimental $A$ value for coronene is, given the uncertainties in the measurements and fitting procedures, very close to the simulated value for single carbon knockout, implying that nearly all of the non-statistical coronene fragments survive in the experiment. In comparison, the values in Table \ref{tab_A} suggest that 35$\%$ and 70$\%$ of anthracene and pyrene fragments survive following the initial knockout.

\begin{figure}
\centering
\includegraphics[width=0.98\columnwidth]{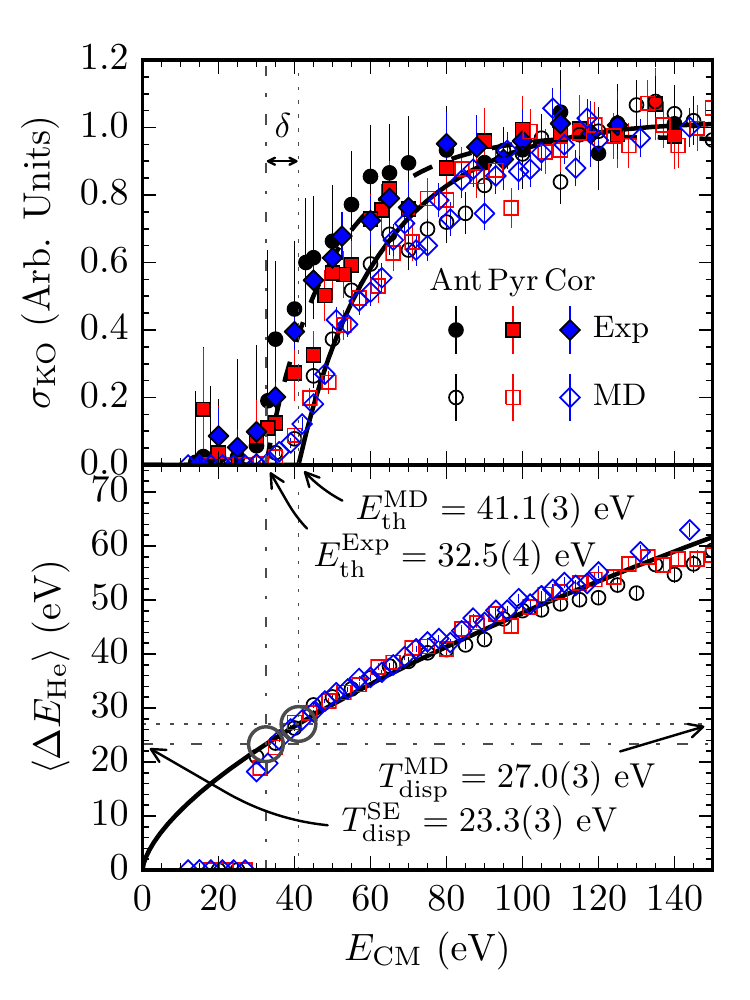}
\caption{Top: Normalized knockout cross sections from the experiments (filled symbols) and the MD simulations (open symbols), with fits to Equation \ref{eq_tao}. Bottom: Average energy lost by the He atom $\Delta E_{He}$ in simulated collisions leading to CH$_x$-knockout. The solid line is a fit to a power law. The displacement energies $T_{disp}$ are the values of $\Delta E_{He}$ at the threshold energy $E_{th}$ and are indicated by rings.}
\label{fig_disp}
\end{figure}

The upper panel of Figure \ref{fig_disp} shows an expanded view of the threshold region for the same combined MD dataset as the lowest panel of Figure \ref{fig_md}, along with the combined, normalized experimental data. Our MD simulations predict a significantly higher threshold energy ($41.0\pm 0.3$ eV), an average offset of $\delta=8.5\pm 0.5$ eV with respect to the experimental threshold ($32.5\pm 0.4$ eV). A likely explanation for this is that the Tersoff potential may give overly strong C-C bonds.


The threshold energies determined by our method are typical values from the fits to Equation \ref{eq_tao}. Specific threshold energies depend on collision angle, impact parameter, and the position of the C atom in the PAH. Indeed, the simulated cross sections contain a low energy tail not accounted for by Equation \ref{eq_tao}, as can be seen in the top panel of Figure \ref{fig_disp}. Another possible contributing effect is vibrations (even zero point vibrations). Meyer \emph{et al.} found zero point vibrations to be essential in describing the electron beam induced knockout of C atoms from graphene \cite{meyer12,meyer13}. However, they deduce the same threshold energy with and without including this effect \cite{meyer13}. 

The lower panel of Figure \ref{fig_disp} shows the mean energy lost by the He projectile, $\Delta E_{He}$, in collisions that lead to knockout of a single carbon atom in our simulations. This energy transfer is well described by a power law for He-PAH$^+$ center-of-mass collision energies between 30 and 150 eV, and the solid line shows a fit yielding $\Delta E_{He}=2.55\times E_{CM}^{0.64}$. We find that this non-linear behaviour is typical for non-zero impact parameter collisions (see the Supporting Information), while zero impact parameter collisions give a linear behavior in agreement with the results reported by Postma \emph{et al.} \cite{postma14}.  The displacement energy, $T_{disp}$, is the value of $\Delta E_{He}$ at the knockout threshold $E_{th}$, and is highlighted by rings in the figure. The value of the displacement energy from the present MD simulations is $T_{disp}^{MD}=27.0\pm 0.3$ eV. We use the fitted value of $\Delta E_{He}$ from the simulations at the experimental threshold energy to determine a semi-empirical value for the displacement energy of $T_{disp}^{SE}=23.3\pm 0.3$ eV. This value is in excellent agreement with the results from electron beam induced carbon knockout from graphene ($T_{disp}=23.6$ eV \cite{meyer13}). It should be noted though that the two experimental situations are different. In the graphene experiments \cite{meyer12,meyer13}, 80-100 keV electrons collide with a graphene monolayer at an angle perpendicular to the plane, while we collide keV PAH cations with individual He atoms essentially at rest in the laboratory system. The latter is equivalent to a situation where He atoms collide with individual PAH cations (at rest) and where the He atoms have well defined energies which can be varied from below 20 eV and up to a few hundred eV. Unlike in the graphene experiments \cite{meyer12,meyer13}, our PAH molecules are randomly oriented when they collide with He. The collision energy required to knock out a carbon atom, the threshold energy, and thus also the displacement energy may be slightly different for different orientations. In addition, PAH molecules contain carbon atoms with slightly different binding energies depending on their positions in the molecules, which also could give some differences in relation to graphene where this is not the case. On the other hand, the the present MD results (which are averaged over all molecular orientations) are in good agreement with our MD results for face-on collisions. This indicates that the face on configuration for PAH molecules is representative of the average. The agreement between our semi-empirical value for the PAH displacement energy and the result for face-on electron collisions on graphene can thus be taken as a strong indication of close similarities between carbon knockout from PAHs and graphene.   


In this Letter, we have reported the first measurements of threshold energies for single carbon knockout (by He) from isolated PAH molecules (anthracene, pyrene, and coronene). The experimental results from PAH$^+$+He collisions are analyzed in view of classical Molecular Dynamics simulations. We arrive at average values of $E_{th}^{Exp}=32.5\pm 0.4$ eV from the experiment and $E_{th}^{MD}=41.0\pm 0.3$ eV from the MD-simulations, with the discrepancy presumably due to shortcomings of the Tersoff potential used to describe the molecular bonds. We derive displacement energies $T_{disp}^{SE}=23.3\pm 0.3$ eV and $T_{disp}^{MD}=27.0\pm 0.3$ eV. We demonstrate that a simple formula (Equation \ref{eq_tao}) may be efficiently used for estimating ion/atom induced absolute carbon knockout cross sections in a wide energy range (10$^1$-10$^5$ eV).  The present results serve as benchmark data for improving theoretical tools and may be used for modelling collision-induced chemical processes involving PAHs. Such processes are believed to be important in  the interstellar medium \cite{micelotta2010,delaunay15}.

This work was supported by the Swedish Research Council (Contract No. 621-2012-3662, 621-2012-3660, and 621-2014-4501). We acknowledge the COST actions CM1204 XUV/X-ray light and fast ions for ultrafast chemistry (XLIC).

\textbf{Supporting Information Available:} Details of the fitting procedure used to extract CH$_x$-loss cross sections at low center-of-mass energies, as well as additional data from our MD simulations, are available in the Supporting Information. This material is available free of charge via the Internet
http://pubs.acs.org.


\section{Experimental and Computational Methods}


Continuous molecular ion beams were produced by means of ElectroSpray Ionization (ESI).  A more detailed description of the ESI setup and beamline is given elsewhere \cite{haag11,stockett14,stockett14b}.  PAH cations were produced using the well established technique \cite{marziarz05} of adding silver nitrate (0.5 mM in methanol) to the solution of the PAH in methanol and/or dichloromethane.  

Following mass selection by a quadrupole mass filter, the molecular ions were accelerated to 1-10 keV and passed through a 4 cm long collision cell containing He gas.  A cylinder lens and two pairs of electrostatic deflector plates serve as a large angular acceptance energy analyzer (fragment mass analyzer).  The fragments are detected on a 40 mm double-stack micro-channel plate (MCP) with a position-sensitive resistive anode.  Absolute total fragmentation cross sections have been determined for each PAH by measuring the attenuation of the primary beam as a function of the pressure in the collision cell. 

To obtain the experimental CH$_x$-loss cross-sections, the relative intensity of the CH$_x$-loss peak is found by integrating the peak and dividing by the integral of all the fragments from which at least one C atom has been lost (fragments where only H atoms have been lost are not included).  The principle source of uncertainty in this procedure is the background subtraction at low energies, where there is also significant blending of the parent ion, CH$_x$-, and C$_2$H$_y$-loss peaks. The relative intensities of the CH$_x$-loss peaks in the experimental spectra are converted to absolute cross sections by multiplying by the absolute total fragmentation cross section obtained by means of the beam attenuation method \cite{stockett14b}.  


For the classical molecular dynamics (MD) simulations we used the reactive Tersoff forcefield \cite{PhysRevB.37.6991,PhysRevB.39.5566} to describe the C-C and C-H bonds using the coefficients for C and H from Ref. \cite{stockett14b}. Interactions between the noble gas atom and each atom in the PAH molecule are modeled using the Ziegler-Biersack-Littmark (ZBL) potential \cite{zbl_pot_book}. The ZBL potential is a screened Coulomb potential used to describe nuclear stopping, the dominant form of energy transfer in collisions in the energy range studied here. In the simulations, a noble gas atom is launched along a randomly selected initial trajectory towards a randomly oriented PAH molecule. The positions and velocities of all atoms are followed for 200 fs using the LAMMPS molecular dynamics software \cite{Plimpton:1995aa}. At the end of the simulation, the structure of the PAH molecule is analyzed for the loss of atoms and the energies of all of the atoms are recorded. This procedure is repeated 10,000 times for each projectile species and collision energy.

\bibliography{carbon_mhs}

\end{document}